\documentclass[twocolumn,preprintnumbers,amsmath,amssymb]{revtex4}

\usepackage{graphicx}
\usepackage{dcolumn}
\usepackage{bm}

\usepackage[latin9]{inputenc}
\usepackage{epstopdf}
\usepackage{bbm}

\usepackage{color}
\definecolor{lightblue}{rgb}{0.2,0.2,0.7}
\definecolor{darkblue}{rgb}{0,0.25,0.5}
\definecolor{redbrown}{rgb}{0.875,0.25,0.125}
\definecolor{darkgreen}{rgb}{0,0.5,0}

\newcommand{\bra}[1]{\ensuremath{\langle #1 \vert}}
\newcommand{\ket}[1]{\ensuremath{\vert #1  \rangle}}

\renewcommand{\b}[1]{\ensuremath{\mathbf{#1}}}
\renewcommand{\H}{\ensuremath{\text{H}}}
\renewcommand{\l}{\ensuremath{\lambda}}

\newcommand{\Tr}{\ensuremath{\text{Tr}}}
\newcommand{\lr}{\ensuremath{\text{lr}}}
\newcommand{\sr}{\ensuremath{\text{sr}}}

\newcommand{\HF}{\ensuremath{\text{HF}}}
\newcommand{\IP}{\ensuremath{\text{IP}}}

\newcommand{\RSH}{\ensuremath{\text{RSH}}}

\DeclareMathOperator{\erf}{erf}

\begin{document}

\title{Adiabatic-connection fluctuation-dissipation density-functional theory based on range separation}

\author{Julien Toulouse$^1$}\email{julien.toulouse@upmc.fr} 
\author{Iann C. Gerber$^2$, Georg Jansen$^3$, Andreas Savin$^1$ and J\'anos G. \'Angy\'an$^4$}\email{janos.angyan@crm2.uhp-nancy.fr}
\affiliation{%
$^1$Laboratoire de Chimie Th\'eorique, UPMC Univ Paris 06 and CNRS, 75005 Paris, France\\
$^2$Universit\'e de Toulouse, INSA-UPS, LPCNO, 31077 Toulouse, France\\
$^3$Fachbereich Chemie, Universit\"at Duisburg-Essen, 45117 Essen, Germany\\
$^4$CRM2, Institut Jean Barriol, Nancy University and CNRS, 54506 Vandoeuvre-l\`{e}s-Nancy, France}


\date{\today}
\begin{abstract}
An adiabatic-connection fluctuation-dissipation theorem approach based on a range separation of electron-electron interactions is proposed. It involves a rigorous combination of short-range density functional and long-range random phase approximations. This method corrects 
several shortcomings of the standard random phase approximation and it is particularly well suited for describing weakly-bound van der Waals systems, as demonstrated on the challenging cases of the dimers Be$_2$ and Ne$_2$.
\end{abstract}

\maketitle
Density functional theory (DFT) is a powerful approach for electronic-structure calculations of molecular and condensed-matter systems~\cite{Koh-RMP-99}. However, one difficulty in its Kohn-Sham (KS) formulation using local density or generalized-gradient approximations (LDA and GGA) is the description of non-local correlation effects, such as those involved in weak van der Waals complexes, bound by London dispersion forces~\cite{DobMcLRubWanGouLeDin-AJC-01}. The adiabatic-connection fluctuation-dissipation theorem (ACFDT) approach is one of the most promising ways of constructing highly non-local correlation functionals. This approach, introduced in wave function theory~\cite{MclBal-RMP-64} and in DFT~\cite{HarJon-JPF-74,LanPer-PRB-77}, consists in extracting non-local ground-state correlations from the linear charge density response function.

Recently, the ACFDT approach has received renewed interest for implementing the random phase approximation (RPA) or other related approximations for atoms, molecules and solids~\cite{Fur-PRB-01,AryMiyTer-PRL-02,FucGon-PRB-02,FurVoo-JCP-05,HarKre-PRB-08,Fur-JCP-08}. The RPA correlation energy is consistent with the use of the exact, self-interaction-free exchange energy. In spite of a number of encouraging results, such as the correct description of dispersion forces at large separation~\cite{DobWanDinMclLe-IJQC-05}, the proper reproduction of cohesive energies and lattice constants of solids~\cite{MiyAryKotSchUsuTer-PRB-02,MarGarRub-PRL-06,HarKre-PRB-08} and an improved description of bond dissociation~\cite{Fur-PRB-01,AryMiyTer-PRL-02,FucNiqGonBur-JCP-05}, several aspects of the RPA are still unsatisfactory.

First, the RPA is a poor approximation to short-range correlations, leading to correlation energies that are far too negative~\cite{YanPerKur-PRB-00}. Second, in a Gaussian localized basis, RPA calculations have a slow convergence with respect to the basis size~\cite{Fur-PRB-01}. Third, the presence of an unphysical maximum (bump) at medium distances in dissociation curves of simple diatomic molecules~\cite{FucNiqGonBur-JCP-05,Fur-PRB-01} indicates an inherent problem which has not yet a fully clarified origin. Fourth, although in principle the orbitals should be calculated self-consistently~\cite{HelBar-PRB-07}, most RPA implementations consist of a post-KS single-iteration calculation, making the choice of the input orbitals sometimes critical. Last but not least, although the main advantage of the RPA is supposed to be the description of dispersion forces, rare gas dimer potential curves calculated from LDA or GGA orbitals are often qualitatively wrong, as shown later.

The poor short-range behavior can be corrected by adding a GGA functional constructed from the difference of the exact and RPA correlation energies of the uniform electron gas~\cite{YanPerKur-PRB-00}, but this so-called RPA+ technique does not lead to consistent improvement~\cite{Fur-PRB-01}. One can go beyond RPA by including exchange-correlation (xc) kernels~\cite{DobWan-PRB-00,FucGon-PRB-02,FurVoo-JCP-05}, but so far it remains imperfect, e.g. local xc kernels produce pair densities that diverge at small interparticle distances~\cite{DobWan-PRB-00,FurVoo-JCP-05}.

In a similar spirit as in the work of Kohn \textit{et al.}~\cite{KohMeiMak-PRL-98}, we propose an ACFDT approach based on a range separation of electron-electron interactions. It involves a rigorous combination of a short-range density functional with one of the possible long-range generalizations of the RPA. The method offers a solution for several of the aforementioned difficulties and is particularly well suited for the description of weakly-bound van der Waals systems.

\vskip 2mm \noindent {\it Theory.} \hskip 2mm
In the range-separated multideterminant extension of the KS scheme, an alternative approach to DFT (see, e.g., Ref.~\onlinecite{TouColSav-PRA-04}), the exact ground-state energy of an electronic system is expressed as 
\begin{equation}
E  = \min_{\Psi} \left\{ \bra{\Psi} \hat{T} + \hat{V}_{ne} + \hat{W}_{ee}^{\lr} \ket{\Psi} + E_{\H xc}^{\sr}[n_{\Psi}]\right\},
\label{EminPsi}
\end{equation}
where $\hat{T}$ is the kinetic energy operator, $\hat{V}_{ne}$ is the nuclei-electron interaction operator, $\hat{W}_{ee}^{\lr} \!=\! (1/2) \iint d\b{r}_1 d\b{r}_2 w_{ee}^{\lr}(r_{12}) \hat{n}_2(\b{r}_1,\b{r}_2)$ is a long-range electron-electron interaction written with $w_{ee}^{\lr}(r)\! =\!\erf(\mu r)/r$ and the pair-density operator $\hat{n}_2(\b{r}_1,\b{r}_2)\! =\! \hat{n}(\b{r}_1)\hat{n}(\b{r}_2)\! -\! \hat{n}(\b{r}_1)\delta(\b{r}_1\!-\!\b{r}_2)$, and $E_{\H xc}^{\sr}[n]$ is the corresponding $\mu$-dependent short-range Hartree-exchange-correlation (Hxc) density functional that Eq.~(\ref{EminPsi}) defines. The minimizing multideterminant wave function, denoted by $\Psi^\lr$, corresponds to a long-range interacting effective Hamiltonian and yields the exact density. The parameter $\mu$ in the error function controls the range of the separation. For $\mu\!=\!0$, the standard KS scheme is recovered: $w_{ee}^{\lr}(r)$ vanishes, $\Psi^\lr$ reduces to the non-interacting KS wave function and $E_{\H xc}^{\sr}[n]$ becomes the usual Hxc functional. For $\mu\!\to\!\infty$, the usual wave function formulation of the electronic problem is retrieved: $w_{ee}^{\lr}(r)$ becomes the full Coulomb interaction, $E_{\H xc}^{\sr}[n]$ vanishes and $\Psi^\lr$ becomes the exact ground-state wave function. For intermediate values of $\mu$, the interaction effects are divided between the long-range interacting wave function $\Psi^\lr$ and the short-range density functional $E_{\H xc}^{\sr}[n]$, and one expects to find better approximations for each piece. Short-range LDA~\cite{TouSavFla-IJQC-04,PazMorGorBac-PRB-06} and several beyond-LDA approximations~\cite{TouColSav-PRA-04,TouColSav-JCP-05,GolWerSto-PCCP-05,FroTouJen-JCP-07} have been proposed for $E_{\H xc}^{\sr}$. Here, we use an ACFDT approach for the long-range part of the calculation. 

In a first step, the minimization in Eq.~(\ref{EminPsi}) is restricted to single-determinant wave functions $\Phi$, leading to a range-separated hybrid (RSH) scheme~\cite{AngGerSavTou-PRA-05} 
\begin{equation}
E_{\RSH}  = \min_{\Phi} \left\{ \bra{\Phi} \hat{T} + \hat{V}_{ne} + \hat{W}_{ee}^{\lr} \ket{\Phi} + E_{\H xc}^{\sr}[n_{\Phi}]\right\},
\label{ERSHminPhi}
\end{equation}
which, in contrast to some range-separated KS schemes~\cite{IikTsuYanHir-JCP-01,GerAng-CPL-05a}, does not include long-range correlation.
The minimizing determinant $\Phi_0$ is given by the self-consistent Euler-Lagrange equation
\begin{equation}
\left( \hat{T} + \hat{V}_{ne} + \hat{V}_{\H x,\HF}^{\lr} +  \hat{V}_{\H xc}^{\sr} \right) \ket{\Phi_0} = {\cal E}_{0} \ket{\Phi_0},
\label{HPhi0}
\end{equation}
where $\hat{V}_{\H x,\HF}^{\lr}$ is a Hartree-Fock (HF) type long-range Hartree-exchange (Hx) potential, $\hat{V}_{\H xc}^{\sr} \!=\! \int d\b{r} \delta E_{\H xc}^{\sr}[n_{\Phi_0}]/\delta n(\b{r}) \hat{n}(\b{r})$ is the short-range local Hxc potential and ${\cal E}_{0}$ is the Lagrange multiplier for the normalization constraint. As usual, $\hat{V}_{\H x,\HF}^{\lr}$ is the sum of a local Hartree part $\hat{V}^{\lr,\mu}_{\H}$ and a non-local exchange part $\hat{V}^{\lr}_{x,\HF}$.

The RSH scheme does not yield the exact energy and density, even with the exact short-range functional $E_{\H xc}^{\sr}$. Nevertheless, the RSH approximation can be used as a reference to express the exact energy as
\begin{eqnarray}
E = E_{\RSH} + E_c^{\lr},
\label{}
\end{eqnarray}
defining the long-range correlation energy $E_c^{\lr}$, for which we will now give an adiabatic connection formula. For that, we introduce the following energy expression with a formal coupling constant $\l$
\begin{eqnarray}
E_{\l}  = \min_{\Psi} \Bigl\{ \bra{\Psi} \hat{T} + \hat{V}_{ne} + \hat{V}_{\H x,\HF}^{\lr} + \l \hat{W}^{\lr}  \ket{\Psi}
\nonumber\\
+ E_{\H xc}^{\sr}[n_{\Psi}] \Bigl\},
\label{ElminPsi}
\end{eqnarray}
where $\Psi$ is a multideterminant wave function, $\hat{W}^{\lr}$ is the long-range fluctuation potential operator
\begin{eqnarray}
\hat{W}^{\lr} = \hat{W}_{ee}^{\lr} - \hat{V}^{\lr}_{\H x,\HF},
\label{Wlr}
\end{eqnarray}
and $E_{\H xc}^{\sr}$ is the previously-defined $\l$-independent short-range Hxc functional.
The minimizing wave function is denoted by $\Psi^{\lr}_{\l}$. For $\l=1$, the physical energy $E=E_{\l=1}$ and density are recovered, as Eq.~(\ref{ElminPsi}) reduces to Eq.~(\ref{EminPsi}). For $\l=0$, the minimizing wave function is the RSH determinant $\Psi^{\lr}_{\l=0} = \Phi_0$. Note that, because the density at $\l=0$ is not exact, the density is supposed to vary along this adiabatic connection.
Taking the derivative of $E_{\l}$ with respect to $\l$, noting that $E_{\l}$ is stationary with respect to $\Psi^{\lr}_{\l}$, and reintegrating between $\l=0$ and $\l=1$ gives
\begin{eqnarray}
E = E_{\l=0} + \int_{0}^{1} d\l \, \, \bra{\Psi^{\lr}_\l} \hat{W}^{\lr}  \ket{\Psi^{\lr}_\l},
\label{}
\end{eqnarray}
with $E_{\l=0}\! =\! \bra{\Phi_0} \hat{T} \!+\! \hat{V}_{ne} \!+\!  \hat{V}_{\H x,\HF}^{\lr} \ket{\Phi_0}\! +\! E_{\H xc}^{\sr}[n_{\Phi_0}]\! =\! E_{\RSH}\! -\!\bra{\Phi_0} \hat{W}^{\lr} \ket{\Phi_0}$. Thus, the long-range correlation energy is
\begin{eqnarray}
E_c^{\lr} = \int_{0}^{1} d\l \Bigl\{ \bra{\Psi^{\lr}_\l} \hat{W}^{\lr}  \ket{\Psi^{\lr}_\l} - \bra{\Phi_0} \hat{W}^{\lr}  \ket{\Phi_0} \Bigl\},
\label{}
\end{eqnarray}
or, using a compact notation,
\begin{eqnarray}
E_c^{\lr} = \frac{1}{2} \int_{0}^{1} d\l \,\, \Tr \left[ w^{\lr} * P_{c,\l}^{\lr} \right],
\label{Eclr}
\end{eqnarray}
where $w^{\lr}$ and $P_{c,\l}^{\lr}$ are four-index representations of the fluctuation potential and the correlation contribution of the two-particle density matrix in a one-electron basis, $*$ stands for contraction of two indices and $\Tr$ is the trace over the remaining two indices. The fluctuation-dissipation theorem is then used to express $P_{c,\l}^{\lr}$ with the imaginary-frequency four-point polarizability $\chi^\lr_\l (iu)$ corresponding to the wave function $\Psi^\lr_\l$ (see, e.g., Ref.~\onlinecite{MclBal-RMP-64})
\begin{equation}
P_{c,\l}^{\lr} = -\frac{1}{2 \pi} \int_{-\infty}^{\infty} du \, e^{-u 0^+} \left[ \chi^{\lr}_{\l} (iu) -\chi_0 (iu) \right] +\Delta^\lr_{\l},
\label{PcfromChi}
\end{equation}
where $\chi_0 (iu)$ is the four-point polarizability for the RSH effective Hamiltonian of Eq.~(\ref{HPhi0}) and $\Delta^\lr_{\l}$ is the contribution coming from the variation of the one-particle density-matrix along the adiabatic connection. The expression of $\Delta^\lr_{\l}$ is straightforward but it is sufficient to write it as $\Delta^\lr_{\l} = F[G^\lr_\l] - F[G_0]$ where $F$ is a known functional, $G^\lr_{\l}$ is the two-point one-particle Green function corresponding to the wave function $\Psi^\lr_\l$ and $G_0$ is the two-point RSH Green function.
Along the adiabatic connection of Eq.~(\ref{ElminPsi}), the Green function $G_{\l}^{\lr}$ satisfies a self-consistent Dyson equation
\begin{eqnarray}
\left(G_{\l}^{\lr} \right)^{-1} = G_{0}^{-1} - \l \left(\Sigma_{\H x}^\lr[G^{\lr}_\l] - \Sigma_{\H x}^\lr[G_0] \right) - \Sigma_{c,\l}^\lr[G^\lr_\l],
\nonumber\\
\label{Dysoneq}
\end{eqnarray}
where $\l \Sigma_{\H x}^\lr$ and $\Sigma_{c,\l}^\lr$ are the Hartree-exchange and correlation self-energies associated with the long-range interaction $w_{ee}^{\lr}$. The long-range polarizability is given by the solution of the Bethe-Salpeter-type equation (see Ref.~\onlinecite{OniReiRub-RMP-02})
\begin{eqnarray}
\left(\chi_{\l}^{\lr} \right)^{-1} = \left(\chi_{\IP,\l}^\lr \right)^{-1} - \l f_{\H x}^\lr  - f_{c,\l}^\lr,
\label{BetheSalpetereq}
\end{eqnarray}
where $\chi_{\IP,\l}^\lr$ is an independent-particle (IP) polarizability whose expression is a frequency convolution of two Green functions $G^{\lr}_\l$, abbreviated as $\chi_{\IP,\l}^\lr\!=\! - i G_{\l}^{\lr} G_{\l}^{\lr}$, and $\l f_{\H x}^\lr\! =\! i \l \delta \Sigma_{\H x}^{\lr}/\delta G_\l^\lr$ and $f_{c,\l}^\lr = i \delta \Sigma_{c,\l}^{\lr}/\delta G_\l^\lr$ are long-range HF-type and correlation kernels.


\begin{figure}[t]
\includegraphics[scale=0.22,angle=-90]{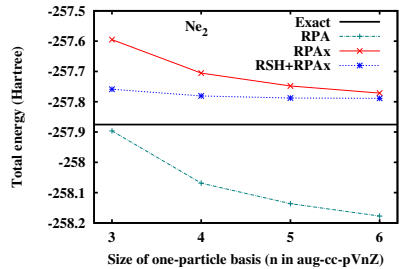}
\caption{(Color online) RPA, RPAx and RSH+RPAx total energies of Ne$_2$ at equilibrium distance (5.84 Bohr) with respect to the basis size.
}
\label{fig:ne2}
\end{figure}

So far, the theory is in principle exact. In the following we introduce the approximation 
\begin{equation}
\Sigma_{c,\l}^\lr\!=\!0,
\label{TheApproximation}
\end{equation}
which corresponds to neglecting long-range correlation in all one-electron properties. Indeed, from Eq.~(\ref{Dysoneq}), this approximation implies that the Green function remains unchanged along the adiabatic connection, i.e. $G_{\l}^{\lr}=G_{0}$ and thus $\Delta^\lr_{\l}\!=\!0$. It also follows that $f_{c,\l}^\lr\!=\!0$ and $\chi_{\IP,\l}^\lr = \chi_{0}$, and Eq.~(\ref{BetheSalpetereq}) then has the structure of the RPA with HF exchange kernel (sometimes refers to as linear response time-dependent Hartree-Fock theory or \textit{full} RPA) that we will designate here by RPAx, by opposition to standard RPA, without exchange kernel (sometimes also called \textit{direct} RPA). In the basis of RSH spatial orbitals, for spin-restricted closed-shell systems, the long-range correlation energy then becomes
\begin{eqnarray}
E_{c}^{\lr} = \frac{1}{2} \int_{0}^{1} d\l \sum_{iajb} \bra{ij}\hat{w}_{ee}^{\lr}\ket{ab} \left( P^{\lr}_{c,\l} \right)_{iajb},
\label{Eclriajb}
\end{eqnarray}
where $ia$ and $jb$ refer to excitations from occupied ($i$ or $j$) to virtual ($a$ or $b$) orbitals, $\bra{ij} \hat{w}_{ee}^{\lr}\ket{ab}$ are the two-electron integrals with long-range interaction, and $(P^{\lr}_{c,\l})_{iajb}$ are the matrix elements of the spin-singlet-adapted $P^{\lr}_{c,\l}$. The one-electron term $\hat{V}^{\lr}_{\H x,\HF}$ in Eq.~(\ref{Wlr}) does not contribute to $E_{c}^{\lr}$ because of the occupied-virtual structure of $P^{\lr}_{c,\l}$. Only singlet excitations contribute to Eq.~(\ref{Eclriajb}), since the two-electron integrals involved vanish for triplet excitations. Note that alternative (but inequivalent) RPAx correlation energy expressions, such as the plasmon formula of Ref.~\onlinecite{MclBal-RMP-64} and the closely related ring CCD approximation of Ref.~\onlinecite{GusHenSor-JCP-08}, require contributions from both singlet and triplet excitations, which may be problematic in systems displaying triplet instabilities, such as Be$_2$. Following the technique proposed by Furche~\cite{Fur-PRB-01}, $P^{\lr}_{c,\l}$ can be obtained as
\begin{eqnarray}
P_{c,\l}^{\lr} &=& 2 \left[ (A_\l-B_\l)^{1/2} M_\l^{-1/2} (A_\l-B_\l)^{1/2} -\b{1} \right], \,\,\,\,\,\,
\label{}
\end{eqnarray}
with $M_\l = (A_\l-B_\l)^{1/2} (A_\l+B_\l) (A_\l-B_\l)^{1/2}$ and the singlet orbital rotation Hessians
\begin{eqnarray}
(A_\l)_{iajb} &=& (\epsilon_{a} - \epsilon_{i}) \delta_{ij} \delta_{ab} 
\nonumber\\
&&+ 2 \l \bra{aj} \hat{w}_{ee}^{\lr}\ket{ib} - \l \bra{aj}\hat{w}_{ee}^{\lr}\ket{bi}, \,\,\,
\label{HessA}
\end{eqnarray}
\begin{eqnarray}
(B_\l)_{iajb} &=& 2 \l \bra{ab}\hat{w}_{ee}^{\lr}\ket{ij} - \l \bra{ab}\hat{w}_{ee}^{\lr}\ket{ji}, \,\,\,
\label{HessB}
\end{eqnarray}
where $\epsilon_{i}$ are the RSH orbital eigenvalues.

This method will be referred to as RSH+RPAx. In the limit of $\mu\!=\!0$, it reduces to the standard KS scheme, while for $\mu \to \infty$ it becomes a full-range ACFDT RPAx approach (with HF orbitals). We note that at second-order in the interaction $w_{ee}^{\lr}$ the RSH+RPAx reduces to the RSH+MP2 method~\cite{AngGerSavTou-PRA-05,GerAng-CPL-05b}.

\vskip 2mm \noindent {\it Computational details.} \hskip 2mm
Equations (\ref{Eclriajb})-(\ref{HessB}) have been implemented in the time-dependent DFT development module~\cite{HesJan-CPL-03} of MOLPRO 2008~\cite{Molproshort-PROG-08}. We perform a self-consistent RSH calculation with the short-range PBE xc functional of Ref.~\onlinecite{GolWerSto-PCCP-05} and add the long-range RPAx correlation energy calculated with RSH orbitals. The range separation parameter is taken at $\mu=0.5$, in agreement with previous studies~\cite{GerAng-CPL-05a}, without trying to fit it. The $\l$-integration in Eq.~(\ref{Eclriajb}) is done by a 7-point Gauss-Legendre quadrature~\cite{Fur-PRB-01}. We use large Dunning basis sets~\cite{Dun-JCP-89} and remove the basis set superposition error (BSSE) by the counterpoise method. The full-range RPA and RPAx calculations have been done with PBE~\cite{PerBurErn-PRL-96} and HF orbitals, respectively. The computational cost of RSH+RPAx is essentially identical to that of full-range ACFDT RPA.

\begin{figure*}[t]
\includegraphics[scale=0.22,angle=-90]{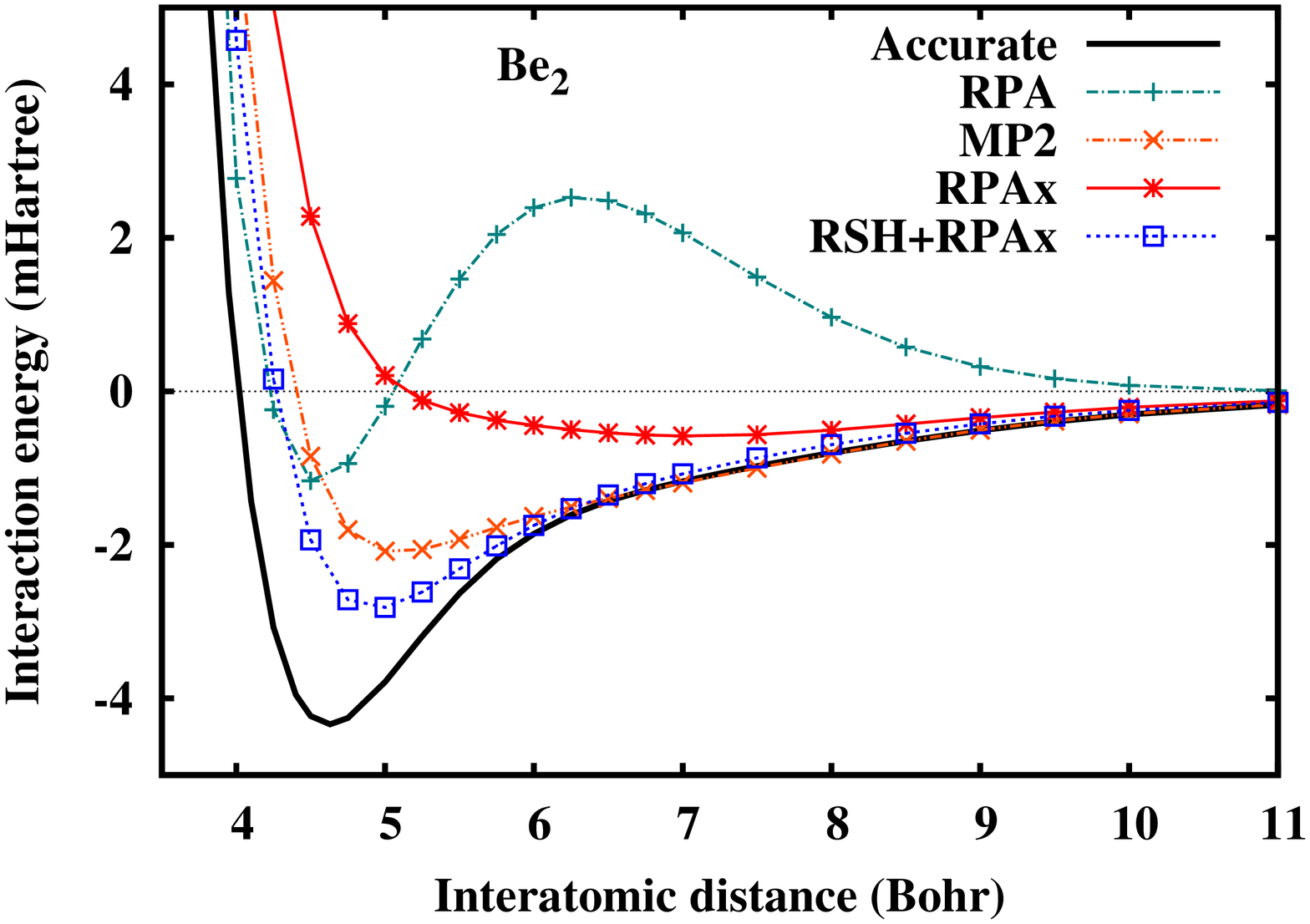}
\includegraphics[scale=0.22,angle=-90]{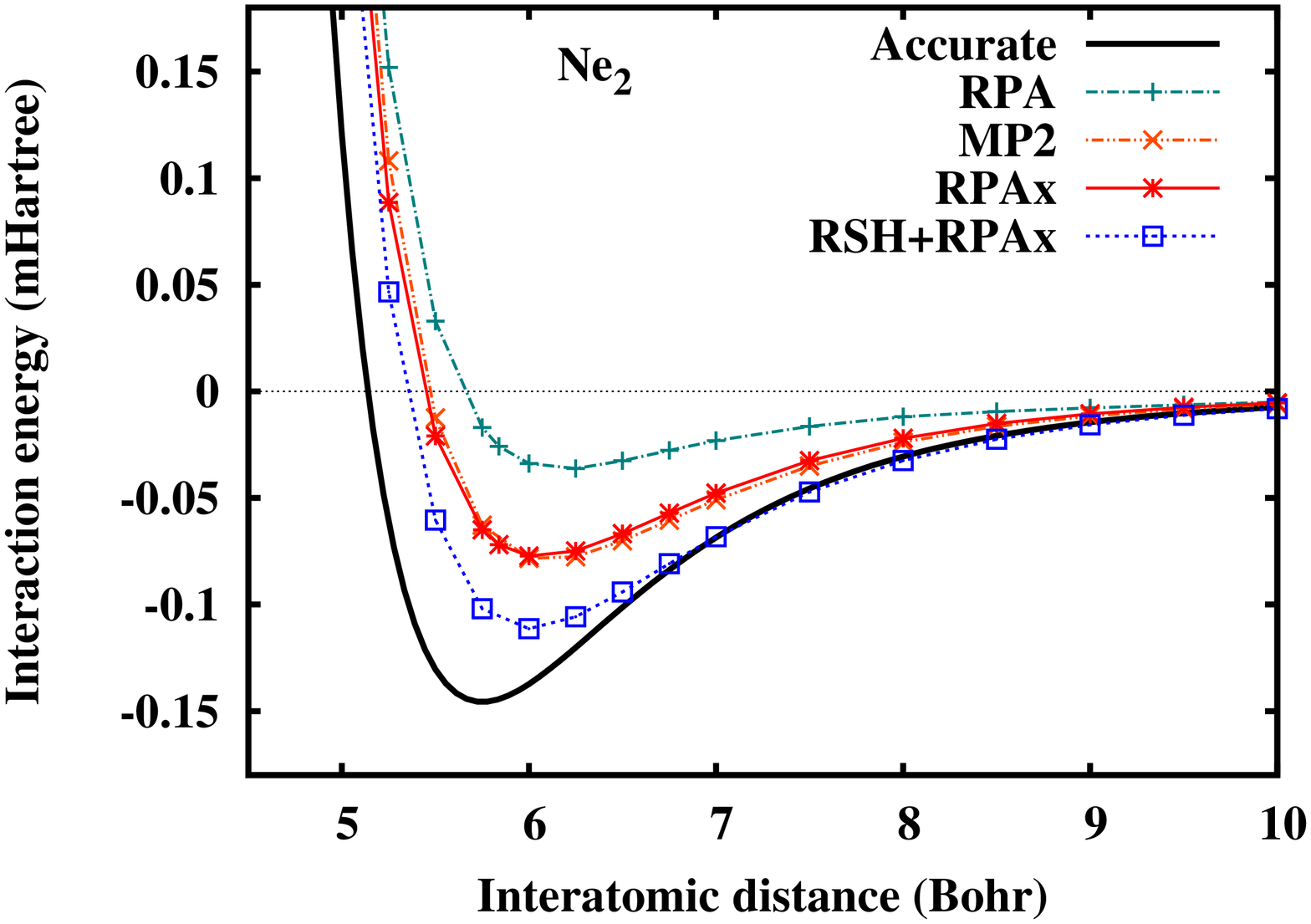}
\caption{(Color online) Interaction energy curves of Be$_2$ and Ne$_2$ calculated in RPA, MP2, RPAx and RSH+RPAx. The basis is cc-pV5Z for Be$_2$ and aug-cc-pV5Z for Ne$_2$. The accurate curves are from Ref.~\onlinecite{RoeVes-IJQC-05} and Ref.~\onlinecite{TanToe-JCP-03}.
}
\label{fig:be2_ne2}
\end{figure*}

\vskip 2mm \noindent {\it Results and discussion.} \hskip 2mm
Figure~\ref{fig:ne2} shows RPA, RPAx and RSH+RPAx total energies with respect to the basis size for Ne$_{2}$. In contrast with full-range RPA and RPAx, a fast convergence is observed for RSH+RPAx, similar to that of standard KS calculations. This improvement is explained by the fact that short-range correlations are compactly described by the short-range density functional. The reduced basis dependence of RSH+RPAx also means smaller BSSE~\cite{TouGerJanSavAng-JJJ-XX-note}. Another important point illustrated by Fig.~\ref{fig:ne2} is that the large RPA overestimation of the total energy, apparent for a large enough basis, is remedied by the RSH+RPAx method thanks to a more accurate description of short-range correlations.

Figure~\ref{fig:be2_ne2} shows the interaction energy curves of Be$_2$ and Ne$_2$. The RPA (with PBE orbitals) fails badly: a large bump for Be$_2$ and an almost completely repulsive curve for Ne$_2$ are observed. A spectacular improvement is obtained with the RSH+RPAx method, which gives physically correct curves, especially accurate at medium and large distances. It also improves over both MP2 and RPAx. It has been verified that the short-range LDA xc functional of Ref.~\onlinecite{TouSavFla-IJQC-04} provides very similar RSH+RPAx interaction curves, indicating a low sensitivity of the method with respect to the short-range functional.

The proposed RSH+RPAx method overcomes many limitations of the RPA or related approaches. The results show that it has the potential to describe successfully weakly-bound van der Waals systems at all distances. It is expected to supersede the RSH+MP2 method~\cite{AngGerSavTou-PRA-05}, especially for systems with small electronic gap. 

\vskip 2mm \noindent {\it Acknowledgements.} \hskip 2mm
We thank F. Furche, J. F. Dobson, T. Gould and G. E. Scuseria for discussions. This work was supported by ANR (07-BLAN-0272).

\bibliographystyle{apsrev}
\bibliography{biblio}

\begin{thebibliography}{38}
\bibitem[{Tou()}]{TouGerJanSavAng-JJJ-XX-note}
\bibinfo{note}{For Ne$_{2}$ at equilibrium distance with aug-cc-pVQZ basis, the
  counterpoise correction on the interaction energy is 16 $\mu$H for RSH+RPAx
  and 62 $\mu$H for standard MP2.}
\end{thebibliography}

\end{document}